\documentclass[numberedappendix]{emulateapj}

\usepackage{multirow}
\usepackage{graphicx}
\usepackage{rotating}
\usepackage{color}
\usepackage{url}

\newcommand{\hz}{\,\mathrm{Hz}}
\newcommand{\project}[1]{\textsl{#1}}
\newcommand{\fermi}{\project{Fermi}}
\newcommand{\rxte}{\project{RXTE}}

\shorttitle{Timing analysis of bursts from magnetars SGR 1806-20 and SGR 1900+14}
\shortauthors{Huppenkothen et al.}

\begin{document}

\title{Quasi-Periodic Oscillations in Short Recurring Bursts of the magnetars SGR 1806-20 and SGR 1900+14 Observed With RXTE}

\author{D. Huppenkothen\altaffilmark{1, 2}, L. M. Heil\altaffilmark{1}, A. L. Watts\altaffilmark{1},  E. G{\"o}{\u g}{\"u}{\c s}\altaffilmark{3}}

\altaffiltext{1}{Anton Pannekoek Institute for Astronomy, University of
  Amsterdam, Postbus 94249, 1090 GE Amsterdam, the Netherlands}
\altaffiltext{2}{Email: D.Huppenkothen@uva.nl}

\altaffiltext{3}{Sabanc\i~University, Orhanl\i-Tuzla, \.Istanbul  34956, Turkey}

\begin{abstract}

Quasi-periodic oscillations (QPOs) observed in the giant flares of magnetars are of particular interest due to their potential to open up
a window into the neutron star interior via neutron star asteroseismology. However, only three giant flares have been observed. We
therefore make use of the much larger data set of shorter, less energetic recurrent bursts. Here, we report on a search for QPOs in a large data set of bursts from
the two most burst-active magnetars, SGR 1806-20 and SGR 1900+14, observed with the Rossi X-ray Timing Explorer (\rxte). We find a single detection in an averaged periodogram 
comprising $30$ bursts from SGR 1806-20, with a frequency of $57\,\hz$ and a width of $5\,\hz$, remarkably similar to a giant flare
QPO observed from SGR 1900+14. This QPO fits naturally within the framework of global magneto-elastic torsional oscillations 
employed to explain the giant flare QPOs. Additionally, we uncover a limit on the applicability of Fourier analysis for light curves 
with low background count rates and strong variability on short timescales. In this regime, standard Fourier methodology and
more sophisticated Fourier analyses fail in equal parts by yielding an unacceptably large number of false positive detections. 
This problem is not straightforward to solve in the Fourier domain. Instead, we show how simulations of light curves can offer
a viable solution for QPO searches in these light curves.

\end{abstract}

\keywords{pulsars: individual (SGR 1806-20, SGR 1900+14), stars: magnetic fields, stars: neutron, X-rays: bursts, methods:statistics}

\section{Introduction}

Neutron stars provide one of the best astrophysical laboratories for the study of nuclear physics under extreme conditions not accessible to standard laboratory experiments: dense, cold, highly asymmetric (neutron-rich) matter up to several times the nuclear saturation density $\rho = 2.8\times10^{14} \, \mathrm{g}\; \mathrm{cm}^{-3}$.
Among the zoo of observable neutron star phenomena, two classes stand out for their peculiar observational properties: Soft Gamma Repeaters (SGRs) and Anomalous X-ray Pulsars (AXPs) \citep[both classes are  \textit{magnetars}; for a general overview, see ][]{woods2006,mereghetti2011}.
They are generally characterised by long spin periods of $2 - 12 \, \mathrm{s}$, a large spin-down derivative and an inferred dipole magnetic field above the quantum-critical limit, $B_{\mathrm{QED}} = 4.4 \times 10^{13} \, \mathrm{G}$ (although in recent years, $3$ sources - out of a total of $26$ sources comprising $21$ confirmed magnetars, and $5$ candidates\footnote{details at \url{http://www.physics.mcgill.ca/~pulsar/magnetar/main.html}} - have been found where the lower limit on the dipole field inferred from spin-down is below $B_{\mathrm{QED}}$,  \citet{vanderhorst2010,esposito2010,rea2010,rea2012,scholz2012,rea2014}). 

Magnetars are of particular interest because of their extensive bursting behaviour across $\sim\!\! 5$ orders of magnitude in duration and nearly $\sim\!\! 9$ orders of magnitude in total isotropic energy. This is especially true for the brightest of their bursting phenomena: giant flares. These vast but short (with durations of $>\! 500\mathrm{s}$) outbursts of hard X-ray emission, with luminosities up to $10^{47}\,\mathrm{erg}\, \mathrm{s}^{-1}$, are rare events believed to occur due to a catastrophic re-structuring of the magnetic field \citep{thompson1995,lyutikov2003}.  The resulting release of energy creates an optically thick pair plasma that slowly radiates the energy away. Analogous to earthquakes, a significant fraction of this energy may also be converted into global oscillations of the star \citep{duncan1998}. These oscillations are of interest to both astrophysicists and nuclear physicists, because if observed, they would provide a unique view into a neutron star's interior (both crust and core). 

The detection of quasi-periodic oscillations (QPOs) in the tails of two giant flares sparked a very active debate about their origin \citep{israel2005,strohmayer2005,strohmayer2006,watts2006}. However, the problem requires complex models \citep[for a general discussion, see ][]{watts2011}: for a full solution, models require inclusion of magnetic fields, both dipole and toroidal components, and a full general relativistic treatment. Additionally, knowledge of the equations of state of both crust and core, but especially the anisotropies in the lower crust, is imperative, as well as inclusion of superfluid and superconducting components. Because we have little understanding of any of these components, models have many degrees of freedom and are highly degenerate. At the same time, giant flares are sufficiently rare that only two out of three observed giant flares have sufficient data to even attempt searches for QPOs, such that the resulting frequencies do not strongly constrain parameter space (for more in-depth discussions of the various models refer to \citealp{samuelsson2007,andersson2009,sotani2007,sotani2008,vanhoven2008,vanhoven2011,vanhoven2012,colaiuda2011,colaiuda2012,gabler2011,gabler2012,gabler2013,passamonti2013a,passamonti2013b,lander2010,lander2011,glampedakis2006,glampedakis2014}).

It seems logical, then, to turn to the giant flares' much smaller cousins, magnetars. These are known to emit short bursts with much less energy, up to $\sim 10^{41} \, \mathrm{erg}$. Unlike the giant flares, they are much more numerous. The data set for the two best-studied magnetars, SGR 1806-20 and SGR 1900+14, spans thousands of such bursts \citep[e.g.][]{gogus1999,gogus2000,prieskorn2012}. It is unclear whether these bursts are smaller manifestations of the underlying physical mechanism that produces giant flares, or a separate phenomenon. If the former is the case, then in principle they might excite star quakes and seismic waves at frequencies similar to those in the giant flares. In this case, they might provide a new avenue for exploring magnetar seismology and constrain theoretical models. Motivated by this hypothesis,  \citet{huppenkothen2014} (using a method developed in \citealt{huppenkothen2013}) studied a sample of 286 bursts from SGR J1550-5418 observed with the Gamma-Ray Burst Monitor (GBM) on board the \fermi\ Gamma-Ray Space Telescope, and found QPOs at $93 \hz$, which is close to the strongest QPO in the giant flare observed from SGR 1806-20, as well as a QPO at $127\hz$ in  periodograms averaged over many bursts. A potential QPO was also found in a single burst, at a much higher frequency of $260 \hz$. The latter was a much broader feature, unlike anything ever seen before in a giant flare.   An earlier search of $152$ individual bursts observed with the Burst and Transient Source Experiment (\project{BATSE}) from several magnetars, using the \textit{Rayleigh} statistic instead of standard periodograms, found only a very marginal detection ($p = 0.01$) in a single burst \cite{kruger2002}. However, this search was restricted to high frequencies due to the effect of the overall burst structure at low frequencies.

Here, we report a search of a similar kind in two bursting episodes of the most burst-active magnetars, SGR 1900+14 and SGR 1806-20. Both have shown giant flares, SGR 1900+14 in 1998 \citep{cline1998,hurley1999,feroci1999} and SGR 1806-20 in 2004 \citep{palmer2005,hurley2004,hurley2005,mazets2005,borkowski2004,mereghetti2005,cameron2005}.  The latter was particularly remarkable as the brightest $\gamma$-ray event ever recorded on Earth, with measurable effects on the terrestrial magnetic field and ionosphere \citep{mandea2006,inan2007}. Both giant flares have shown QPOs at frequencies between $18\hz$ and $1840\hz$ at energies between $2 \, \mathrm{keV}$ and $200 \, \mathrm{keV}$ \citep{strohmayer2005,israel2005,strohmayer2006,watts2006}, and have a rich data set of short bursts.
Here, we focus on two burst episodes observed with the \project{Rossi} X-ray Timing Explorer (\rxte) in 1996 (SGR 1806-20) \citep{gogus2000} and 1998 (SGR 1900+14) \citep{gogus1999} \footnote{Note that \citet{elmezeini2010} searched a subset of the SGR 1806-20 data set considered here for QPOs. However, flaws in the data analysis procedure as described in the Appendix of \citet{huppenkothen2013} render the QPOs discovered in this analysis invalid.}. 

In Section \ref{sec:data} of this paper, we briefly describe the data and data processing procedures. In Section \ref{sec:analysis}, we review the statistical methodology of searching for QPOs in Fourier-transformed light curves, which is used in the rest of this paper, and we report on the results of its application, both for individual bursts as well as averaged periodograms from larger burst samples. Subsequently, we show one limit where the applied method unexpectedly failed and characterise that failure via extensive simulations, in Section \ref{sec:weakbursts}, before describing an alternative way to identify and characterise the significance of potential detection in Section \ref{sec:dnest}. We conclude with a discussion of the theoretical implication of our results in Section \ref{sec:discussion}.

\section{Data}
\label{sec:data}

We employed burst data collected from the two strongest-field magnetars, SGR 1806-20 and SGR 1900+14,  with the Proportional Counter Array (PCA) on board \rxte. SGR 1806-20 was observed during an active period in 1996 (observation IDs 20165 and 10223) and SGR 1900+14 during an active period in 1998 (observation ID 30410). These active periods, a subset of the thousands of bursts observed from both magnetars, were chosen both for the large number of bursts within a relatively short time interval (such that we can easily average consecutive bursts and search for long-lived as well as re-excited QPOs), and for the quality of the observations; all five detector units (PCUs) were in operation for most of the bursts, which allows us to detect even weak ones.

We include  $558$ bursts from SGR 1806-20 and $229$ bursts from SGR 1900+14, all investigated in \citet{gogus2001}. These bursts were bright enough to allow their $T_{90}$ durations (i.e., the time around the peak count rate in which 90\% of all photons arrive at the detector) to be measured \citep{gogus2001}. We accumulated burst data starting from the $T_{90}$ start times and lasting for the course of their $T_{90}$ durations. PCA data for these two magnetars were collected in GoodXenon and Event modes. Events were extracted from channels covering the $2-60\,\mathrm{keV}$ energy range at the intrinsic bin size provided by the observation mode, which is 1 $\mu$s for Good Xenon mode and 125 $\mu$s for the Event modes.  

Because our analysis makes extensive use of Fourier methods (see Section \ref{sec:analysis} below for details), instrumental effects that change the distribution of arriving photons need to be taken into account in the data analysis.
Following \citet{zhang1995} and \citet{jahoda2006}, we correct the Fourier-transformed burst periodograms for dead time effects. Dead time occurs when the X-ray detector is momentarily unresponsive after a photon impinges on it. In \rxte, there are two main types of dead time: (1) dead time after arrival of a photon, where the channel in which the photon arrived is paralysed for $10\mu\mathrm{s}$, and (2) dead time after a very large event (VLE), a photon with an energy much higher than the dynamic range of the detector, which saturates the amplifier. The latter paralyses the detector for $170\mu\mathrm{s}$. While both effects operate on very short timescales, much shorter than the timescales of interest here, the resulting loss of photons modifies the distribution of photon arrivals away from a Poisson distribution, and consequently also modifies the distribution of powers in the periodogram. Note that dead time depends very strongly on count rate: the brighter a source, the stronger the effect on the periodogram. Thus, dead time corrections are especially important for the brightest bursts, however, since the effects become appreciable even at moderate count rates of $\sim 2000 \,\mathrm{counts}\,\mathrm{s}^{-1}$, virtually all bursts need to be corrected. We use equations (10) and (13) of \citet{jahoda2006} to correct for dead time. The corrections are defined per PCU, whereas we use light curves combined from all active units in our analysis. Thus, the given normalisation constants are incorrect; we fit for these constants using a Maximum Likelihood approach, and correct for the resulting deviation in both noise level and periodogram shape.

\section{Periodogram Searches}
\label{sec:analysis}

Magnetar bursts require special care when performing Fourier analysis on their light curves. Because they have, by their very nature as bursts, a start and an end, they are non-stationary processes. Note that stationarity does not imply a constant light curve: it merely implies that the average properties of the mean and variance in the light curve over any given time interval must be the same as over any other interval of the same length (as opposed to, for example, a light curve with an overall trend). Non-stationarity leads to deviations in the statistical distributions and the shape of the power spectrum (defined as the square of the Fourier amplitudes, for an introduction see \citealp{vanderklis1989}), such that standard methods are not easily applicable.
Here, we use the Bayesian periodogram methods described in \citet{huppenkothen2013} to deal with the effects of non-stationarity at low frequencies. In short, we compute the periodogram of a light curve with a high time resolution, here $dt = 0.5/2048 = 2.44 \times 10^{-4} \, \mathrm{s}$, which allows us to search up to a Nyquist frequency of $\nu_{\mathrm{Nyquist}} = 2048 \, \mathrm{Hz}$. For light curves that obey stationarity over the timescales of interest, standard Fourier methodology applies, and the statistical distributions of the resulting power spectra are well known. The bursty nature of our light curves introduces high variance at long timescales; correspondingly the periodogram shows high power at low frequencies. We model this power with an empirical function. Experience has shown that simple or broken power laws can model a large range of burst phenomena \citep{huppenkothen2013}. 

Consequently, we perform two tasks: (1) a model selection task, to ascertain whether the periodogram may be represented by a simple power law or whether it requires a more complex model and (2) a QPO search task, where we compare the maximum powers of a large number of simulations to the maximum power after dividing out the best-fit broadband model in the observed periodogram. For the model selection task, we fit the periodogram with both a simple and a broken power law and compute the likelihood ratio. We then sample from the posterior distribution of the simpler model via Markov Chain Monte Carlo \citep[MCMC; using the freely available \textit{Python} code \project{emcee},][]{foreman2013}, and simulate periodograms from draws of that posterior distribution. These periodograms are again fit with both models, such that we can build a distribution of likelihood ratios for realisations of the simpler model. This allows us to compute a posterior p-value, such that we can accept or reject the simple model. 

In the second step, we draw from the posterior distribution of the model chosen in the model selection step, again via MCMC, and create a large number of simulated periodograms from these draws. We fit each periodogram with the preferred model, and find the highest data/model outlier. We can then compare the distribution of data/model outliers as derived from the simulations of broadband noise only with the highest data/model outlier in the observed periodogram. If the observed value is very unlikely given the p-value derived from these simulations, one may say with relative confidence that we have detected a QPO at the frequency of the highest data/model outlier in the data. Note that while this approach automatically corrects for the fact that we have searched over a broad range of frequencies, we still need to correct for the fact that we also have searched over a large number of bursts: the more frequencies or bursts one searches, the more likely it becomes to see an outlier purely by chance. 

The analysis presented above makes a strong assumption about the data: our choice of a $\chi^2$-distributed likelihood around the model power spectrum implies that the periodogram is the result of a pure, stationary noise process. This is not strictly true, but as shown in \citet{huppenkothen2013}, it is a conservative assumption that holds for all but the lowest frequencies in the periodogram. At high enough frequencies, where the shape of the periodogram is effectively hidden by noise, the method becomes equivalent to the standard tests against a $\chi^2$ distribution with two degrees of freedom for an unbinned periodogram, as described for example in \citet{vanderklis1989}.

For details on the analysis procedure, including extensive simulations on simulated bursts, as well as the limitations of the method, see \citet{huppenkothen2013} and \citet{vaughan2010}.
\begin{figure*}[htbp]
\begin{center}
\includegraphics[width=18cm]{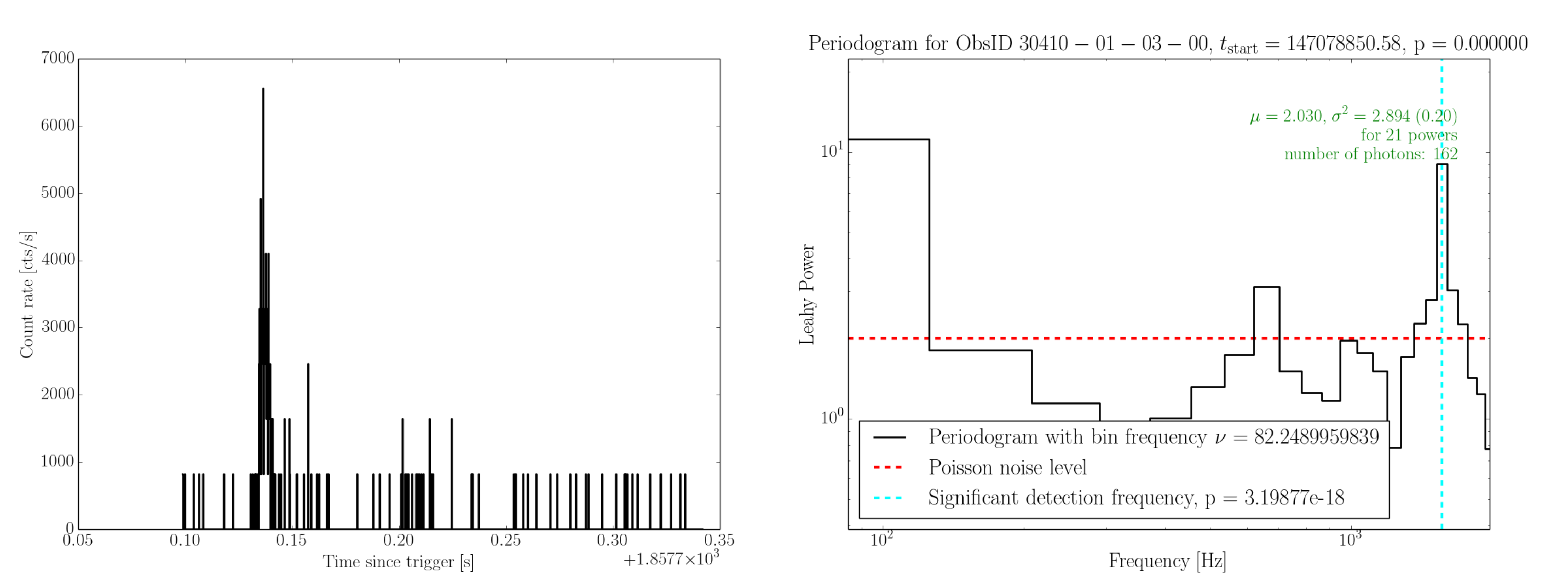}
\caption{Light curve (left) and binned periodogram (right) of a single burst observed from SGR 1900+14. The burst has very few photons ($N_{\mathrm{photons}} = 162$), and is representative of the sample. The periodogram shows a strong, frequency-dependent modulation across the entire frequency range, with the strongest signal at $1560\hz$ with a very high significance ($p = 3.19 \times 10^{-18}$, single trial). Because the periodogram deviates strongly from the distributions we test against, standard tests as well as the method from \citet{huppenkothen2013} potentially overestimate the significance. This will be explored in detail in Section \ref{sec:weakbursts}.}
\label{fig:psd_individual_example}
\end{center}
\end{figure*}

\subsection{Individual Burst Search}
\label{sec:psd_individual}

We search a total of $558$ bursts from SGR 1900+14, and $229$ bursts observed with SGR 1806-20, excluding all bursts that saturated the detector or showed drop-outs in the light curve. 
Each periodogram was corrected for dead time as explained in Section \ref{sec:data}. 
For each burst, we first constructed a distribution of the likelihood ratio for a power law versus a broken power law model from $1000$ simulated periodograms using the simpler model, and
chose the broken power law to model the burst periodogram if the power law model was rejected at the $p < 0.05$ level. This is fairly conservative by design: we prefer to overfit a simple burst rather than underfit a periodogram 
with complex structure, which may then be mis-attributed to a QPO feature. We could have simply chosen the broken power law only and avoided this step; however, in order to be consistent with previous analyses \citep{huppenkothen2013,huppenkothen2014} as well as to preserve the capability of characterising the broadband variability properties, we kept this step as part of the analysis.
Subsequently, the selected broadband noise model was used to simulate $1000$ periodograms and build the
distribution of data/model outliers, which we used to test for significant QPOs across the $50 - 2000 \hz$ frequency range. Frequencies below $50 \, \mathrm{Hz}$ are discarded, because on these timescales the overall structure of
the burst likely dominates the periodogram, and the broadband noise model no longer provides an adequate representation of the data. Any burst with a probability of $p < 5 \times 10^{-3}$ (see below for a justification of this particular limit) in at least two different frequency bins
for observing the recorded maximum power under the assumption of pure noise is said to contain a candidate detection. For candidate QPOs with frequencies $> 250 \hz$, above which the distribution of powers
should converge to the classically expected $\chi^2$ distribution with two degrees of freedom, we compute the classical p-value \citep{groth1975}, and thus avoid having to run large numbers of simulations, which 
would quickly become prohibitively computationally expensive. 
We search both the unbinned periodogram, and binned periodograms at $14$ different frequency resolutions. The frequency resolutions we choose are integer multiples of the native periodogram frequency, and thus the actual frequency resolution changes with each periodogram as they are of different lengths. We space bin frequencies logarithmically (between $3$ and $300$ times the original frequency resolution), such that we achieve a reasonable coverage of the entire frequency range. This ensures that we are sensitive not only to QPO signals with widths smaller or approximately equal to the unbinned frequency resolution, but also broader signals without having to perform a model selection for the presence of a QPO component in the model. For each of the binned periodograms, we can then extract the maximum data/model outlier in the same way as for the unbinned periodogram. We bin the simulated periodograms in the same way as the data, such that we can construct posterior distributions for the maximum data/model outlier at each bin frequency and search for QPOs at each frequency resolution.

We refine the sample of candidate detections using this classical p-value for high-frequency candidates. Our initial p-value threshold of $p < 5 \times 10^{-3}$ is not very constraining, given that we search nearly $800$ bursts and 
across $14$ different frequency resolutions. While the number of frequencies within a periodogram is automatically taken into account by the design of the method, the number of individual bursts and frequency resolutions
searched is not. Thus, the p-value needs to be corrected in order to reflect the correct probability of observing a given event by chance. We adjust the threshold on the classical p-value for all candidates with frequencies
$> 250 \hz$ such that only detections corresponding to a $4\sigma$ threshold ($p < 5.7 \times 10^{-9}$ for a single trial, or $p < 6.33 \times 10^{-5}$ taking into account all bursts and frequency resolutions) remain
as candidates. Note that this latter procedure \textit{only} concerns detections with frequencies above $> 250 \hz$, not detections with frequencies below.

We find 15 candidate QPO detections in SGR 1900+14, and 15 candidates in SGR 1806+20 that meet our criteria for a QPO detection, with frequencies between $160\hz$ and $1900\hz$. While our algorithm
flagged these features as significant and QPO-like, there is a clear flaw in the analysis method: an examination of the burst periodograms reveals that the powers are not $\chi^2$ distributed even at high frequencies,
a result confirmed by small probabilities when comparing the distributions of powers above $250 \hz$ with the theoretically expected distribution through a Kolmogorov-Smirnov test (see Figure \ref{fig:psd_individual_example} for an example for how the powers deviate from the expected mean up to high frequencies). This implies that the comparison we are making between the data and the assumed probability model is not fair, or, in other words, standard tests as well as the method from \citet{huppenkothen2013} will potentially overestimate the significance.

It is not immediately clear what causes these irregularities in the periodogram at high frequencies. On the one hand, this is the frequency range where the spectrum should be dominated entirely by Poisson statistics,
as was indeed the case for the bursts from SGR J1550-5418 observed with \fermi/GBM. On the other hand, these frequencies are still too low for dead time effects, described in more detail in Section \ref{sec:data}, to have an
appreciable effect on the shape of the periodogram. It thus seems that there must be an intrinsic property of the bursts that leads to the observed deviations from the expected shape. This possibility will be further
explored in Section \ref{sec:weakbursts}; an alternative QPO search on the bursts in question will be described in more detail in Section \ref{sec:dnest}.

\begin{figure}[htbp]
\begin{center}
\includegraphics[width=9cm]{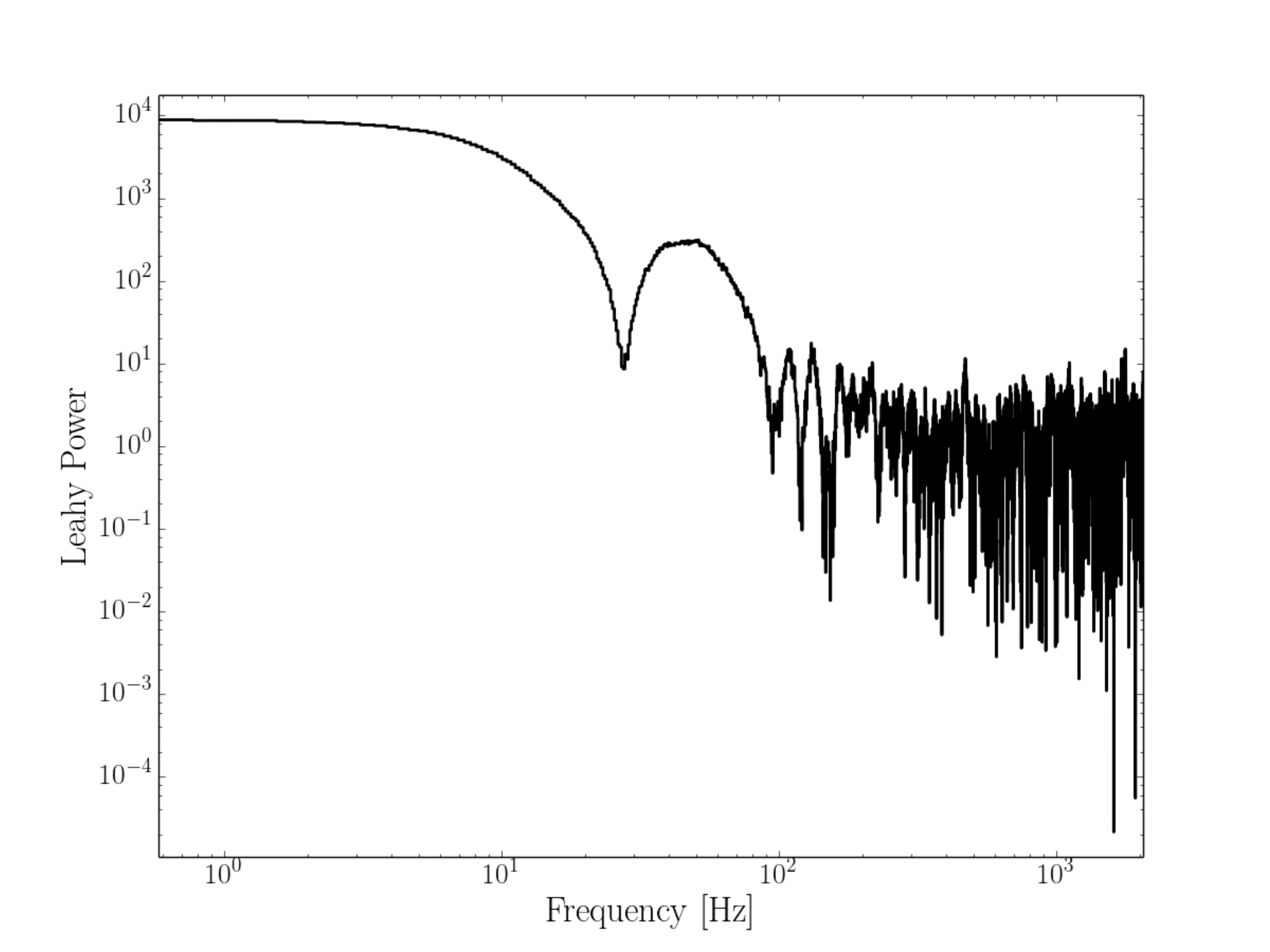}
\caption{Periodogram of an example burst from SGR 1900+14. The powers below $100 \hz$ clearly do not follow a $\chi^2$ distribution with $2$ degrees of freedom around the underlying power spectrum, as is the assumption for our analysis method, nor are neighbouring frequencies independent. The shape of the periodogram at low frequencies is the hallmark signature of a strong burst envelope---the overall shape of the burst---dominating the periodogram. We exclude bursts such as these from the averaged periodograms searched in Section \ref{sec:psd_average}, since aberrant power spectral shapes like the strong feature at $\sim 60 \hz$ are not captured by our model, and can potentially dominate the averaged periodogram even when many bursts are included in the average.}
\label{fig:envelope_example}
\end{center}
\end{figure}

\subsection{Averaged Periodograms}
\label{sec:psd_average}
We construct averaged periodograms from bursts that are close together in time, in order to test the hypothesis that a QPO could persist for hundreds of seconds, or else be re-excited in consecutive bursts at a comparable frequency. Additionally, averaging periodograms from different bursts can drastically increase the signal-to-noise ratio, if a signal persists across bursts. We compute waiting times between burst start times, i.e. the time interval between consecutive bursts. For bursts separated by a gap in the data (this can be either the time between consecutive observations, or a detector drop-out), this time interval is very long. All bursts with a waiting time of less than $500$ seconds between consecutive bursts are then grouped together in clusters. This number is chosen such that we do not create stretches that cross observations, while also creating clusters of bursts large enough to allow for averaging. Within each cluster, we pick the burst with the largest burst $T_{90}$ duration, and construct light curves for all bursts in the cluster with that duration. This allows us to create periodograms with the same number of frequencies, which are easier to average. For the following analysis, we choose all clusters with at least 30 bursts for SGR 1900+14, and all clusters with at least 20 bursts for SGR 1806-20, to account for the intrinsically lower number of bursts in the latter sample, while preserving a high signal-to-noise ratio for both. All periodograms are Leahy-normalised before averaging. This implies that at low frequencies, where the burst variability introduces large Fourier amplitudes, there will be a deviation from the expected statistical distributions at each frequency if the bursts differ substantially in flux.

\begin{figure*}[htbp]
\begin{center}
\includegraphics[width=\textwidth]{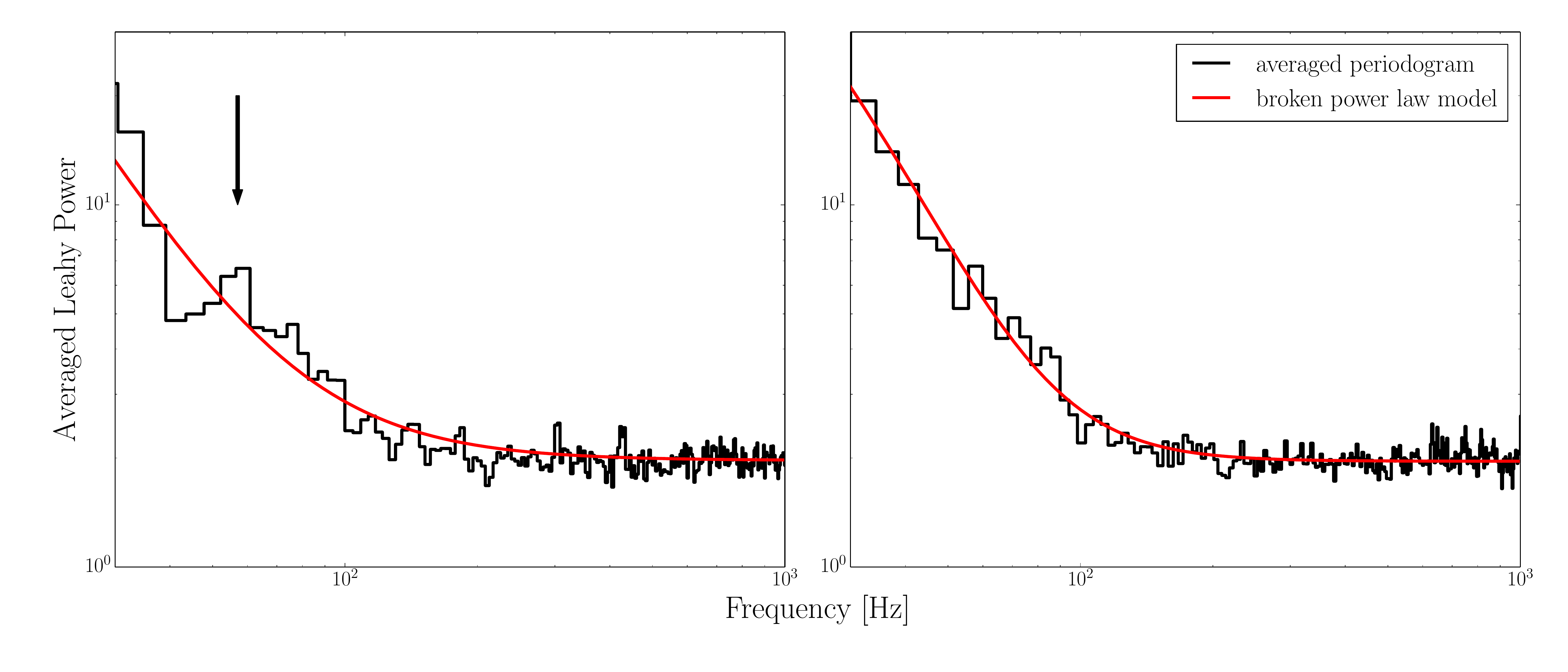}
\caption{Average periodogram with candidate detection (left, black arrow) listed in Table \ref{tab:psd_avg_results} and without detection (right). We show the periodogram averaged to $4 \,\hz$ (black) together with the maximum a posteriori estimate of
the broken power law model (red). The number of averaged periodograms ($m=30$ and $m=23$) is roughly comparable.}
\label{fig:psavg_comp}
\end{center}
\end{figure*}

\begin{figure}[htbp]
\begin{center}
\includegraphics[width=9cm]{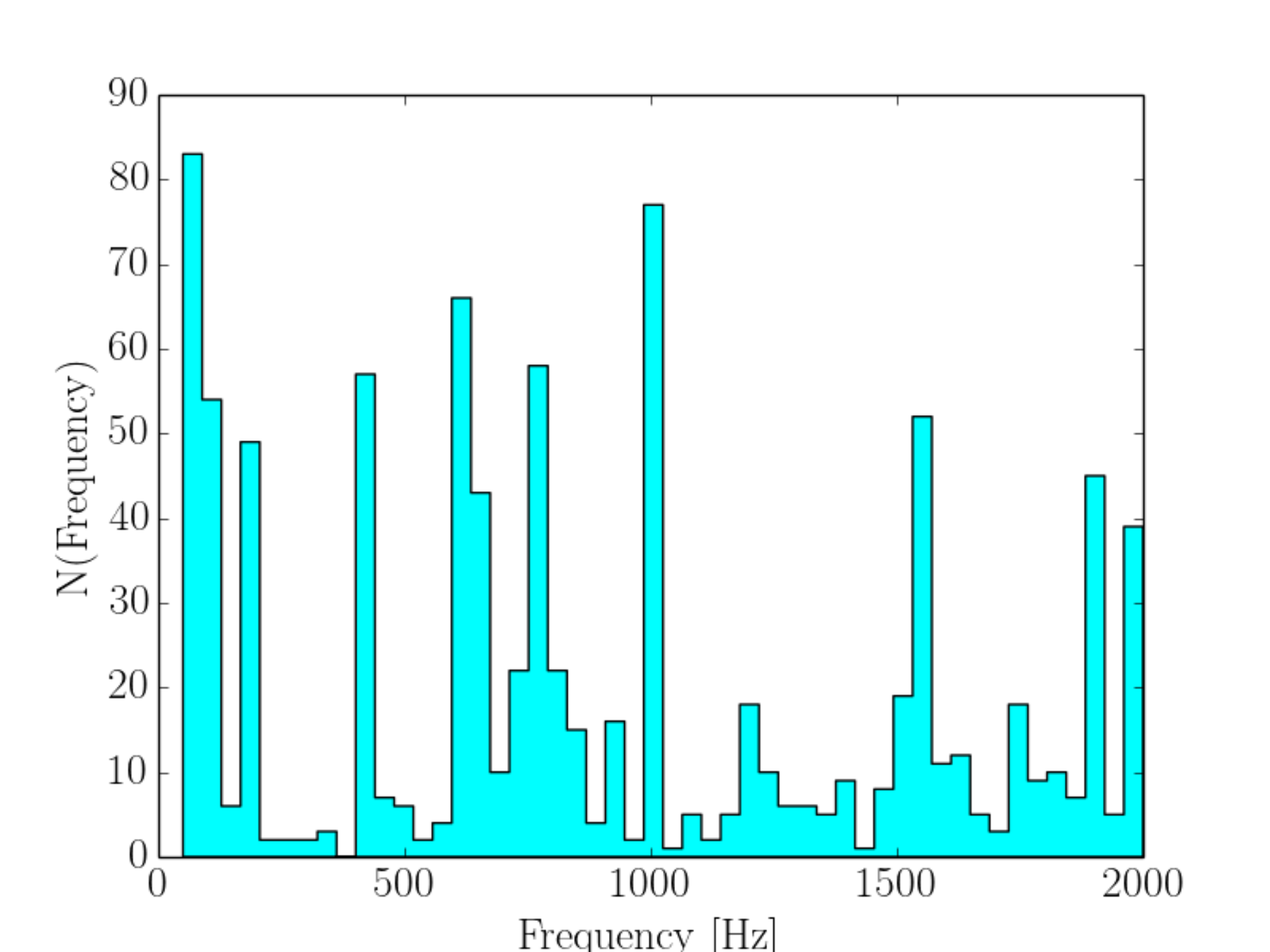}
\caption{Distribution of frequencies of maximum powers for $1000$ averaged periodograms from randomly sampled sets of $30$ bursts each for SGR 1806-20. The averaged periodograms were fit with a broken power law, and the maximum power as well as the frequency of that power extracted from the data/model residuals. It is clear that the distribution of frequencies is not uniform: there are high peaks at frequencies where excess power preferentially occurs.  Here, we show an example where the frequency was averaged to $\sim 2 \hz$, however, the observed distributions are stable across a large range in frequency resolutions.}
\label{fig:psd_avg_freqdist}
\end{center}
\end{figure}
For SGR 1900+14, we create 15 clusters in this way, containing between $6$ and $69$ bursts each. These clusters have durations (from the first burst in the cluster to the last) between $775$ and $3257\,\mathrm{s}$, longer than the instrument-imposed maximum duration of a cluster of $330 \, \mathrm{s}$ for our previous analysis of \fermi/GBM data of SGR J1550-5418 \citep{huppenkothen2014}. Eight of these clusters have more than $30$ bursts, which we subsequently combine to produce eight periodograms, with the smallest sample averaged over $38$ bursts and the largest averaged over $75$ bursts.
For SGR 1806-20, we create 19 clusters with between 1 and 44 bursts and a total duration in each cluster $17$ and $3242$ seconds. We extract five clusters with between $20$ and $44$ bursts each. Note that the lower number of bursts averaged for SGR 1806-20 compared with SGR 1900+14 leads to a lower sensitivity for QPO detections in the former data set, as the inclusion of more bursts results in a higher signal-to-noise ratio in the final averaged periodogram.

Our Bayesian QPO search algorithm finds candidate detections in six averaged periodograms from SGR 1900+14 and in four in averaged periodograms from SGR 1806-20, at frequencies between $50 \hz$, the lower boundary of our search, and $1900 \hz$. 
Two effects may strongly affect the probability of detecting a QPO in averaged periodograms.

First, because of the deviations from the expected $\chi^2$ distribution described in Section \ref{sec:psd_individual} as well as effects of the differences in burst flux on the statistical distributions of powers in the averaged periodogram, we need to test whether the averaged periodograms could be dominated by power from a single burst, which would lead us to draw wrong conclusions about the averaged periodogram. For low-frequency candidate signals below $100 \hz$, we screen the periodograms of all individual bursts used to produce the averaged periodograms, and exclude those that are clearly dominated by the overall burst process below $100 \hz$ (see Figure \ref{fig:envelope_example} for an example). For high-frequency detections, we search the results of the single-burst QPO analysis for detections at the relevant frequencies in the bursts that make up the averaged periodograms, and exclude those where detections were found. 

Second, even after exclusion of single bursts with strong features that might affect the results of the averaged periodograms, it is possible that the averaged periodograms are affected overall by the non-$\chi^2$-distributed features described in Section \ref{sec:psd_individual}. While some effects may cancel out when averaging many individual bursts together, we cannot assume so a priori. 

In order to test the robustness of the remaining signals, we create averaged periodograms from random samples of bursts from each magnetar. For each averaged periodogram, we create $1000$ random samples of bursts from either SGR 1900+14 or SGR 1806-20, with the same number of bursts averaged as for the averaged periodogram in question. The simulations created this way are not entirely statistically independent: for 1000 simulations, and between 20 and 80 bursts per averaged periodogram, individual bursts will be part of several simulations. Using these simulated periodograms, we can only test the hypothesis that a QPO could be long-lived, or re-excited in bursts that are temporally close, but we cannot test against the null hypothesis that there is no QPO: it is possible that a potential QPO signal could be randomly excited at the same frequency in many bursts, irrespective of whether they occur close together in time or not. In this case, a signal would appear insignificant with respect to the simulations, whereas it is simply present in many bursts.

At the same time, these simulations also allow us to test whether there could be problems with the underlying power spectrum of averaging many bursts. For our key assumption, a $\chi^2$-distributed random variable to hold the maximum powers in each averaged periodogram derived from random samples of bursts should be uniformly distributed across the entire frequency regime. If this is not true and if in fact, many averaged periodograms cluster at specific frequencies, this could indicate problems with the underlying assumption which might be evidence against a QPO at a given frequency.
\begin{deluxetable*}{lcccccc|ccccc}
\label{tab:avgrms}
\tablewidth{500pt}
\tablecolumns{6}
\tablecaption{QPO Detections in Averaged Periodograms}
\tablehead{
\colhead{Observation ID} &
\colhead{$N_{\mathrm{bursts}}$} &
\colhead{$t_0$ [MET]}&
\colhead{min $T_{90}$ [s]} &
\colhead{max $T_{90}$ [s]} &
\colhead{$\nu_0$ [Hz]} &
\colhead{$\Delta\nu$ [Hz]} &
\colhead{posterior} &
\colhead{simulated}\\
\colhead{} &
\colhead{} &
\colhead{} &
\colhead{} &
\colhead{} &
\colhead{} &
\colhead{}&
\colhead{$p$-value} &
\colhead{$p$-value}}
 \startdata
 10223-01-03-010 	&	30	&	90907122.0225 	& 	0.064	&	4.84		&	57	&	4.4	&	$<10^{-4}$	&	$10^{-3}$ \\
 %%% CONTINUE TABLE
 
 \enddata
 \tablecomments{This table summarises the properties of the single credible QPO detection emerging from the averaged periodogram of SGR 1806-20. }
\label{tab:psd_avg_results}
\end{deluxetable*}

After testing both those periodograms with the most prominent burst profiles taken out, as well as testing against distributions of randomly sampled bursts, we find only one significant signal remains: in an average of $30$ bursts observed from SGR 1806-20, we find a significant detection (posterior p-value $p < 10^{-4}$, from random samples of bursts $p = 10^{-3}$, corrected for the number of frequencies searched) at $57 \hz$, with an estimated width of $4.4 \hz$ (see also Figure \ref{fig:psavg_comp}). This QPO is at a frequency where many averaged periodograms from individually averaged bursts show their maximum power as well. While this signal is not an outlier with respect to the frequency distribution, it is an outlier in terms of its power, and is thus more likely to be due to an actual QPO. There is no remaining significant QPO in any of the averaged periodograms from SGR 1900+14.

We note that all other potential QPOs initially flagged as significant by our Bayesian algorithm are not significant when compared to randomly sampled bursts. We also note that the distribution of frequencies of maximum powers extracted from averaged periodograms of randomly sampled bursts is highly non-uniform (see Figure \ref{fig:psd_avg_freqdist} for an example): there are well-defined peaks in the frequency distribution. For SGR 1806-20, the three highest peaks are at $50 - 90 \hz$, $980 - 1020 \hz$ and $1550 - 1590 \hz$; for SGR 1900+14, the peaks are at $80 - 120 \hz$, $1140 - 1180 \hz$, and $1400 - 1440 \hz$. It is unclear what underlying process creates these non-uniform distributions. There could, of course, be QPOs at these frequencies that are continuously excited and re-excited. However, low-frequency features in particular are very sensitive to the broadband noise model: at these frequencies, the power spectrum of the burst itself often supplies significant amounts of power, distorting both the shape and the statistical distributions of the resulting periodogram. Similarly, we have shown in Section \ref{sec:psd_individual} that the statistical distributions of powers are not statistically distributed following the expected $\chi^2$ distribution in the individual burst periodograms, even at high frequencies, and that neighbouring frequencies are often correlated. It is thus possible that the averaged periodograms show an accumulated version of these irregularities. The frequency distribution would then be reminiscent of relevant timescales in the burst, which need not necessarily be related to periodic or quasi-periodic processes. With current methods, it is impossible to distinguish between those two alternatives.

In conclusion, we find only one credible candidate QPO in SGR 1806-20 that is an outlier both with respect to the theoretically expected distributions for an averaged periodogram and with respect to randomly sampled bursts (see Table \ref{tab:psd_avg_results} and Figure \ref{fig:psavg_comp} for details).

\section{Burst Periodograms in the Low-Noise Limit}
\label{sec:weakbursts}

The observed deviations of the high-frequency powers in individual bursts from the expected statistical distributions pose an important problem for and a strong limitation on QPO searches in magnetar bursts with \rxte. In order to perform QPO searches with any degree of confidence, we need to understand the underlying cause of the observed distributions, and find a way to mitigate its effects on the periodogram. In the following, we explore the causes for the observed deviations using simulations of magnetar bursts and characterise the changes in the periodogram based on these simulations. 

While observations of magnetar bursts with \rxte\ suffer from less noise than those made with \fermi/GBM, the integrated number of photon counts over a burst is a factor of $10$ lower than for GBM. This means we are searching for QPOs in the limit of low photon counts, which can have an appreciable effect on the overall statistics, and lead to a deviation from the expected statistical distributions even at high frequencies.  Searching for periodic and quasi-periodic signals in low-count rate data is not a new problem: at high energies, especially in $\gamma$-ray astronomy, a number of statistical tests for detecting periodicities exist even when the light curve consists of few photons \citep[e.g.][]{buccheri1983,dejager1988}. However, these methods focus exclusively on the detection of periodic signals against a constant or at the very least stationary background and are thus biased when used on burst light curves such as those considered here. \citet{kruger2002} employed the Rayleigh statistic, a commonly used test for periodicity in photon counting data that requires no binning, to search for periodic signals in a large number of both magnetar bursts and GRBs observed with BATSE. They found no significant evidence of a QPO, however, they worked in a regime with much higher count rates, and restricted themselves to signals $400\,\hz$ to avoid contamination by low-frequency variability. To our knowledge, there has been no systematic study of the effects of low count rates and low background on the power spectrum of a non-stationary light curve. We thus explore the regime where this deviation becomes important via simulations of light curves with low photon counts, of both simple flat Poisson noise and bursts.
The overall simulation strategy is as follows:

(1) For a given total number of photons, we compute the expected number of counts per time bin. 
(2) We simulate $n_{\mathrm{sim}} = 10000$ light curves from the computed count rate either by picking from a Poisson distribution with a mean equal to the count rate for each time bin, or by normalising a burst shape such that the integrated number of photons will be distributed around the expected number of counts. For low count rates, this will result in a large number of bins with no photons. The integrated number of photons in each light curve will not be $N_{\mathrm{tot}}$ exactly, but fall on a distribution around that value.
(3) For each simulated light curve, we create the periodogram and pick the maximum of the resulting powers above $1000 \hz$. The high cut-off frequency ensures that we do not accidentally include any of the low-frequency, power-law-like variability in our estimates. We then bin the periodogram at different bin factors representative of those chosen for the SGR burst light curves ($b = [5, 10, 20, 50]$). Again, from each periodogram, we pick the highest power above $1000 \hz$.
(4) In order to compare the distribution of maximum powers with theoretical predictions, we simulate the same number of powers as in the unbinned and binned periodograms created in (3) from a $\chi^2$ distribution with $2$ degrees of freedom, $P \sim \chi^2_2$, as expected for periodograms of pure white noise (a flat, Poisson-distributed light curve). For the binned periodograms, the powers are still distributed as a $\chi^2$ distribution, now with $2b$ degrees of freedom, and scaled by $b$: $P \sim \chi^2_{2b}/b$.
(5) Finally, we compare the resulting distributions of maximum powers from the theoretically expected distributions and the distributions of maximum powers from the periodograms derived from simulated light curves by computing the $99\%$ upper quantile of the distribution, and comparing this to the $99\%$ upper quantile expected for a $\chi^2$ distribution. Ideally, the difference between those two quantiles should be zero. For positive differences, the distribution of maximum powers is shifted toward higher powers for the simulations, resulting in likely spurious detections. For negative values, the distribution of maximum powers from the simulations of light curves is shifted toward lower powers compared to a $\chi^2$ distribution, potentially resulting in missed QPO detections.

\subsection{Flat Light Curves}
\label{sec:analysis_lcsims}
As a first step, we simulate simple constant light curves with characteristics similar to the observed bursts: short duration ($T_{90} < 1 \, \mathrm{s}$), high time resolution ($dt = 0.5/2048 \, \mathrm{s} = 2.44\times 10^{-4} \, \mathrm{s}$) and low numbers of photons (between 100 and 10000 photons per burst). 
We produce a large number of simulated light curves for different values of the total number of photons per light curve, in order to test how a low photon count rate affects the periodogram. We space the total photon count $N_{\mathrm{tot}}$ logarithmically, and simulate for $N_{\mathrm{tot}} = [100, 200, 500, 1000, 2000, 5000, 10000]$, keeping all other parameters (e.g. burst duration and time resolution) the same.

For flat light curves, the resulting distributions are close to a $\chi^2$ distribution with two degrees of freedom, and remain this way even for low photon counts. The difference in $99\%$ quantiles between the simulated powers and the expected distribution range from $-0.7$ to $+0.25$, with the difference asymptotically approaching $0$ when averaging neighbouring frequency bins.
This indicates that a few photons alone are not enough to make the resulting periodogram deviate significantly from the expected distribution.

\subsection{Simulated Burst Light Curves}
\label{sec:weakburstsims}

Since a low number of photons alone does not explain the observed deviations from the theoretically expected distribution, we instead simulate simple, single-peaked bursts similar to those observed from SGR 1806-20 and SGR 1900+14 with \rxte.
Simulating a burst adds additional parameters to the model. We model a burst as a single spike of the form

\begin{equation}
\phi(t) = A \left\{\begin{array}{ll}\exp(t/\sigma) & \mbox{for $t<t_\mathrm{max}$}\\ \exp{-t/(\sigma s)} & \mbox{for $t\geq t_{\mathrm{max}}$}\end{array}\right. \, ,
\label{eqn:spikemodel}
\end{equation}

\begin{figure}[htbp]
\begin{center}
\includegraphics[width=9cm]{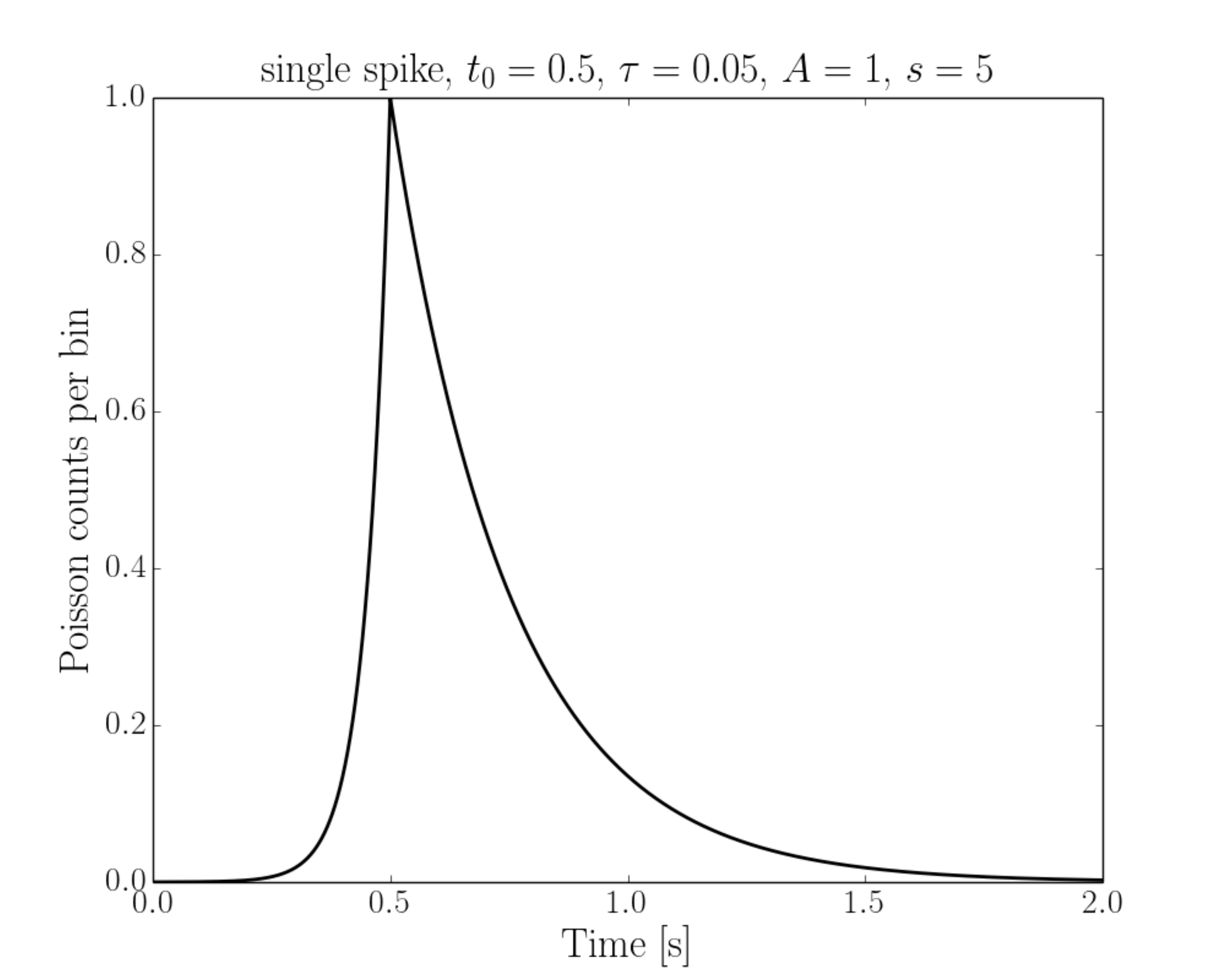}
\caption{Example of a single component of our spike model, defined in Equation \ref{eqn:spikemodel}. The defining parameters are the amplitude $A$ in counts per bin, the position $t_0$ of the peak of the spike,  the exponential rise timescale $\tau$, and the exponential fall timescale, parametrised by a skewness parameters $s$, such that the exponential fall timescale becomes $s\tau$. We model magnetar bursts as a linear combination of these shapes, plus a global parameter accounting for the (flat) background count rate.}
\label{fig:example_spike}
\end{center}
\end{figure}

where $A$ is the amplitude of a spike, $\sigma$ the rise time, $t_\mathrm{max}$ the location in time of the spike maximum, and $s$ a skewness parameter that sets how the decay time is stretched ($s > 1$) or contracted ($s < 1$) compared to the rise time (see Figure \ref{fig:example_spike} for an example of the model). 
For our exploratory analysis here, we restrict ourselves to testing the effect of three parameters in a single-spiked burst: a sharp rise or drop in the light curve (parametrised by varying the rise time of the burst), a change in amplitude, and a change in background count rate. For each combination of rise time, amplitude and 
background count rate, we simulate $n_{\mathrm{sim}} = 10000$  light curves by picking from a Poisson distribution, as in step (2) above, and repeat steps (3) to (5) for these simulations as well.

\begin{figure*}[htbp]
\begin{center}
\includegraphics[width=18cm]{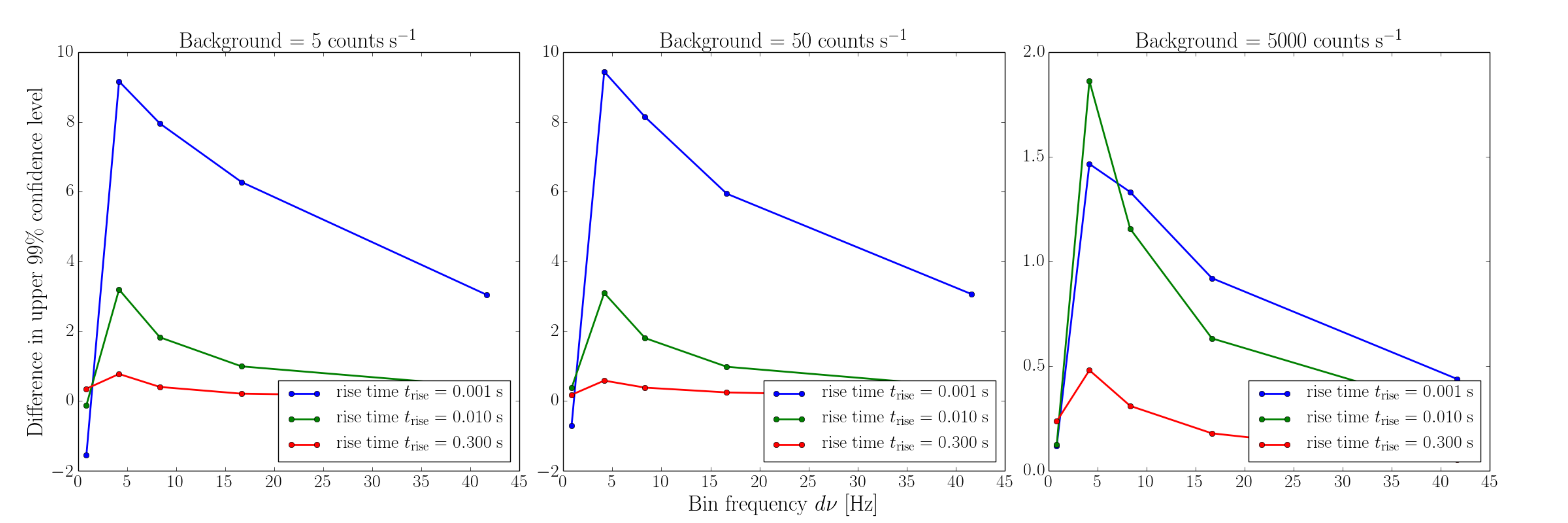}
\caption{Deviations of the distribution of powers $>1000 \hz$ in simulated bursts as a function of frequency binning, burst rise time, and background count rate. For three background count rates ($5 \, \mathrm{counts} \; \mathrm{s}^{-1}$ (left), $50 \, \mathrm{counts} \; \mathrm{s}^{-1}$ (middle) and $500 \, \mathrm{counts} \; \mathrm{s}^{-1}$ (right)) we plot the difference in the $99\%$ quantile from distributions of simulated powers derived from 10000 simulated burst periodograms for three different rise times, vs.~ the $99\%$ quantile of the theoretically expected distribution, as a function of the binning frequency. This difference provides an estimate of how likely we are to over- or underestimate the significance of a given power when compared to the theoretically expected distribution. For positive differences, the observed maximum power in a periodogram is derived from a distribution effectively shifted to the right of the theoretically expected distribution: we are likely to overestimate the significance of that maximum power. Conversely, a negative difference implies a shift of the distribution of powers to the left compared to the theoretically expected distribution, thus we are more likely to underestimate the significance of the observed maximum power in a given burst periodogram. The deviation of the $99\%$ quantile from the theoretical expectation depends strongly on both the rise time, which effectively sets the smallest timescales with power in the periodogram, and the background count rate. Note that we varied the amplitude of the burst as well, but omit a comparison between amplitudes here. A higher amplitude exacerbates the effect for bursts with a low background and a sharp rise.}
\label{fig:weak_bursts}
\end{center}
\end{figure*}

We vary the rise time from $0.001 \, \mathrm{s}$ to $0.03 \, \mathrm{s}$. This is generally shorter than the rise times inferred for bursts in these two sources based on the time between the start point of the $T_{90}$ interval to the time of maximum count rate. However, we note that here we are not interested in the total rise time of a burst (which may have multiple peaks), but in the rise time of each individual peak, which may play a crucial role in determining the frequency up to which power is observed in the periodogram, and motivates our choice for the range of rise timescales simulated. 

We vary the the background counts from $0.001$ counts per bin (corresponding to a background count rate of $\approx 5 \, \mathrm{counts}\, \mathrm{s}^{-1}$) to $10$ counts per bin (corresponding to a count rate of $\approx 5 \times 10^{4}\, \mathrm{counts}\, \mathrm{s}^{-1}$). The background for the PCA detector on board of \rxte\ is approximately $20\, \mathrm{counts}\; \mathrm{s}^{-1}$ per detector, thus well within the range of simulated values.

We find that the deviations from the theoretically expected statistical distribution of powers in many burst periodograms arise from a combination of factors (see Figure \ref{fig:weak_bursts} for an illustration). The low background conspires with sharp rises to create visible features even at high frequencies. This is unsurprising: the sharper the rise, the shorter are the timescales that the Fourier transform decomposes. Correspondingly, the strongest effects are observed for a short rise time, $t_{\mathrm{rise}} = 10^{-3}\,\mathrm{s}$. The powers at high frequency become correlated, and one of the primary assumptions in our analysis---statistical independence of neighbouring frequencies---is broken. This leads to broader distributions of powers, especially when binning over neighbouring frequencies. In a data stream with significant background photon counts, the resulting deterministic structures in the power spectrum are hidden underneath the noise. For \rxte\ data, this is not true: even above $1000\hz$, the periodogram is dominated by structures that arise from the burst itself, even when that burst is a simple, single-peaked structure without any QPO-like features. The effect is strongest for bursts with the weakest background and the sharpest rise times. Our simulations indicate that an increase in amplitude exacerbates the effect: a brighter burst, for the same rise time, automatically implies a sharper rise, thus increasing the power at high frequencies. The effect almost always shifts the distribution of powers to a higher power, and comparisons of observed powers with the theoretically expected statistical distribution will be biased toward overestimating the significance of the observed signal. Using the method from \citet{huppenkothen2013} it is thus more likely to make false positive errors and claim significance for a feature that is not, in fact, a QPO.

We note that our QPO search of bursts from magnetar SGR J1550-5418 observed with \fermi/GBM did not suffer from these problems. In part, this is due to the generally higher sensitivity of the instrument, leading to an increase of a factor of 10 in count rates. Additionally, the background in \fermi/GBM is higher than for \rxte, with $\sim 320 \, \mathrm{counts} \; \mathrm{s}^{-1}$ per detector in the $50 - 300 \, \mathrm{keV}$ energy range \citep{meegan2009}. This ensures that the periodogram at high frequencies follows the expected statistical distribution, and makes QPO searches using models of the periodogram feasible. Beyond that, it is possible that there are intrinsic differences between the two burst samples. Perhaps bursts from the two magnetars considered here have intrinsically shorter rise times. It is also possible that this is an energy-dependent effect: \fermi/GBM observes at a higher energy range than \rxte. Detailed modelling of rise times as a function of energy would be necessary to determine whether this effect is indeed energy-dependent, something that is beyond the scope of this work.
\label{sec:dnest}

\section{Burst Periodograms from Models of the Light Curve: Simulating Candidate Detections}

The results of Section \ref{sec:weakburstsims} make it clear that for short transient events observed with \rxte, the main assumption of the method used in \citet{huppenkothen2013} no longer holds. Even at high frequencies, the powers in the periodogram do not follow a $\chi^2$ distribution with two degrees of freedom around the underlying power spectral model. The simulations also show that we are far more likely to overestimate the significance of a signal due to a sharp rise and a low background than to underestimate the significance. This is a problem that cannot easily be solved in the Fourier domain. Instead, the most straightforward way would be to model the burst light curves directly, and compare the periodogram of the observed data to the periodograms of realisations of the model. This, however, presents us with a new set of problems: there is no simple, straightforward way to model magnetar bursts. Indeed, the variety of shapes in the temporal domain originally prompted us in \citet{huppenkothen2013} to consider power spectral models instead. However, in order to understand whether any of the candidate detections in Section \ref{sec:psd_individual} are real, simulations of light curves are essential. 

In order to simulate the light curves of bursts with candidate detections, we require two ingredients: (1) a simple, yet flexible model that can effectively encompass the large range of burst shapes observed in the data and (2) an algorithm that can efficiently traverse parameter space for the model we consider, and return samples from high-probability regions of that parameter space without too much human intervention. Below, we give a brief outline of a new method that satisfies both requirements, which will be described in more detail in a forthcoming paper.

We model the light curve using the model defined in Equation (\ref{eqn:spikemodel}) as a superposition of individual spikes. We use a Poisson likelihood and hierarchical priors on the parameters to construct a Bayesian model of the light curve, where the components in the superposition are of the type described in Equation \ref{eqn:spikemodel}. The model explicitly includes the number of components $N$ as a parameter to be inferred with the model parameters themselves. Each model component has four parameters (peak time, exponential rise timescale, skewness and amplitude). Another parameter models the background count rate (assumed to be constant over the interval of the burst). Priors are largely uninformative and either exponential \citep[amplitude, rise timescale; see also][]{skilling1998}, uniform (position of component peak between start and end time of a burst; equal probability of skewness in either direction; number of components), or log-uniform (hyperparameters for the exponential priors on amplitude and rise timescale, background parameter). 

We then use trans-dimensional MCMC sampling \citep[in the form of diffusive nested sampling; ][]{brewer2011} to sample the posterior distribution of parameters, including the number of components. Overall, the algorithm reliably finds both positions and shape parameters for the components modelling the brightest spikes. There is a degree of ambiguity for weaker features on whether there should be a component, but this ambiguity is generally small. A different choice of prior for amplitude and exponential rise time reveals some sensitivity to the prior for the weakest features: a log-normal prior tends to include more components with amplitudes close to the background. For our purposes here, this is of little importance for two reasons. First, because the background is so low in the \rxte\ data, there will be little ambiguity over the presence of a spike in the data. Additionally, the resulting light curve, be it composed of a superposition of low-amplitude spikes and a flat background or a simple flat background alone, will appear the same in Fourier space. The details of the method as well as an application to magnetar bursts will be reported in a forthcoming paper (D.~ Huppenkothen et al, in preparation). 

From draws of this distribution, we simulate light curves using the appropriate Poisson statistics to account for the effects of photon counting, and then create periodograms out of these light curves. These periodograms can then be directly compared to the periodogram of the observed data, such that we can create posterior p-values in much the same way as we have done for the periodogram simulations in \citep{huppenkothen2013}. 

\begin{figure}[htbp]
\begin{center}
\includegraphics[width=9cm]{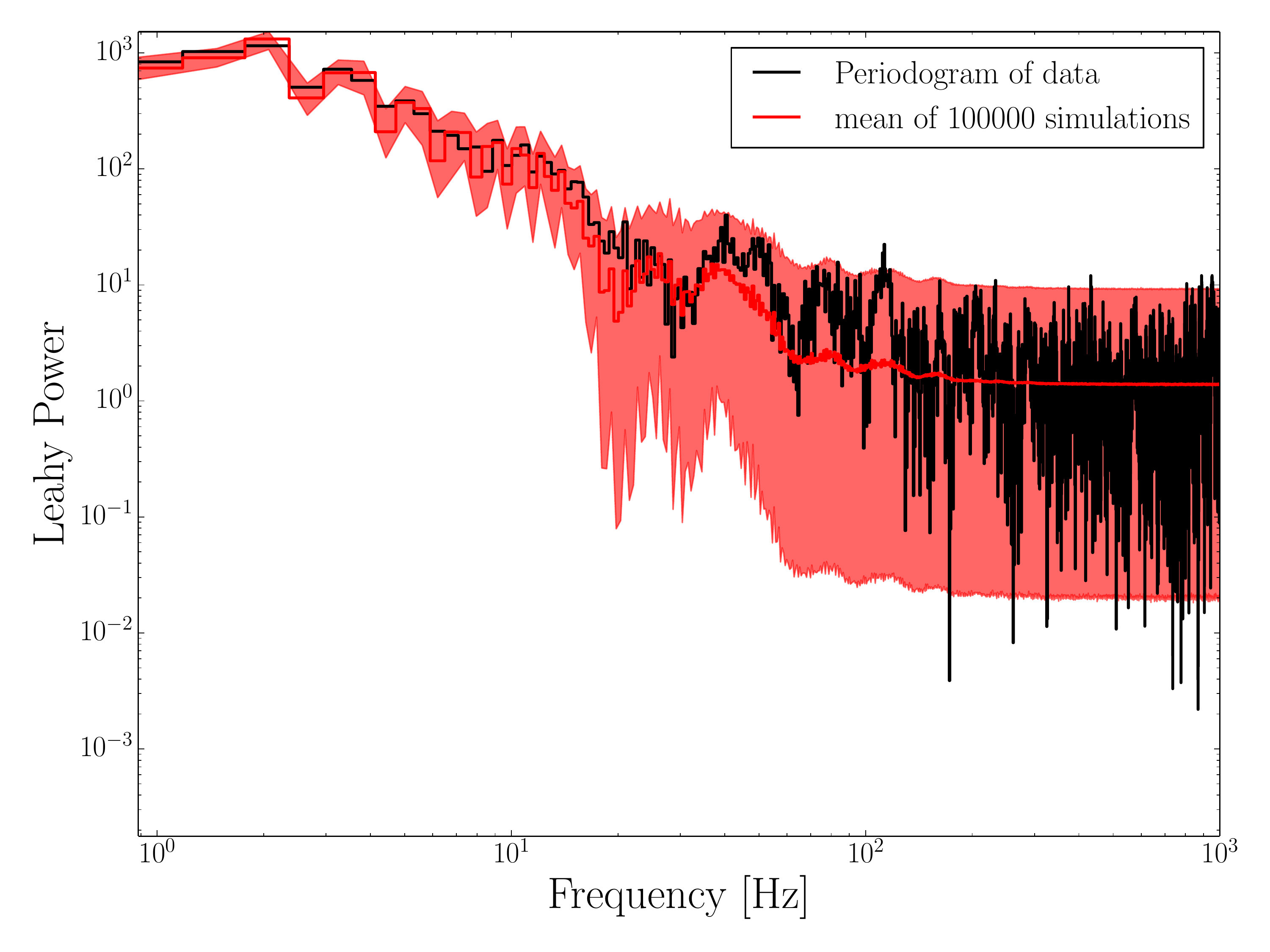}
\caption{Example of a burst periodogram (black) and periodograms of simulated light curves, produced as described in the text. We show the mean out of $10^5$ simulations as the red line; the shaded area encloses the $1\%$ and $99\%$ quantiles. At low frequencies, the irregular shape
of the periodogram due to the overall burst morphology is evident, as are the deviations from the expected (much wider) distribution.  It is reproduced well by the simulations.}
\label{fig:burst_lcmodel_ps}
\end{center}
\end{figure}
For each candidate burst, we create $10^6$ artificial light curves and, consequently, $10^6$ simulated periodograms. We fit each observed periodogram and simulated periodogram with a broken power law, and extract the maximum data/model outlier from both unbinned and binned periodograms. We restrict ourselves to the broken power law model for both the data and simulations, because the full model selection procedure would be too computationally expensive to run on each simulated periodogram. Choosing the more complex model in this case ensures that we are conservative; we are more likely to overfit the spectrum, and thus overfit a potential low-frequency, broad QPO feature, than to underfit a broadband noise feature which will subsequently be falsely detected as a QPO. Since our initial search has not unearthed any low-frequency features that could easily be mistaken for QPOs by the broadband noise model, this is a safe choice.

An example periodogram is shown in Figure \ref{fig:burst_lcmodel_ps}. We show both the periodogram of the burst itself and contours of the $1\%$ and $99\%$ quantiles of the $10^6$ simulated periodograms. Particularly at low frequencies, the deviations from the expected distributions are clearly visible; the shape of the periodogram at these frequencies is complex, with many features that could be mistaken for QPOs, and the distributions are very narrow, indicating that a deterministic process dominates this part of the spectrum. At high frequencies, the spectrum evens out on average, but individual realisations of the burst may still have---and in fact do have---significant deviations from the expected (flat) power spectrum.

We apply this method to all $30$ candidate detections described in Section \ref{sec:psd_individual} in order to confirm or reject the presence of any QPOs in individual magnetar burst periodograms. We compute distributions for the maximum data/model outlier for various binning factors, in order to be able to test whether an observed power is an outlier compared to this distribution, and compute the posterior p-value that this might be the case for each periodogram and binning factor. 
We find four very marginal QPOs, three in bursts from SGR 1806-20, one in a burst from SGR 1900+14. However, none of these candidates have a p-value $p < 10^{-3}$. Given the number of bursts searched, we cannot consider any of these candidates significant detections.

\section{Discussion}
\label{sec:discussion}

We searched for QPOs in a data set comprising magnetar bursts from active periods of the two strongest-field magnetars, SGR 1806-20 and SGR 1900+14. These sources present a particularly interesting case, because giant flares with QPOs present in the tails have been observed from both of them.
We find a candidate detection at $57 \,\hz$ in an averaged periodogram of 30 consecutive bursts observed from SGR 1806-20, with a significance of $10^{-3}$. The total energy in this stack of bursts is $\sim 10^{39} \, \mathrm{erg}$ (order of magnitude estimate, with burst energies taken from \citealt{gogus2000}). Interestingly, this QPO is close in both frequency and width to a QPO observed in the 1998 giant flare from SGR 1900+14, which had a frequency of $53\,\hz$ and a width of $5\,\hz$, and is also similar in width and amplitude, although not in frequency, to those seen in an analysis of \fermi/GBM data of SGR J1550-5418 \citep{huppenkothen2014}. Although it is somewhat surprising that it is this frequency that appears in averaged bursts from SGR 1806-20, instead of the strongest giant flare QPO at $92\,\hz$ that was also detected in SGR J1550-5418, the signal's frequency nevertheless fits naturally within the framework generally employed to explain the QPOs in giant flares. The most plausible explanation for the frequencies in the giant flares, which lie in the range $18 - 1840 \, \hz$, is that they represent global seismic oscillations of the star.   Within the context of current models, a frequency of $57 \,\hz$ would be a relatively low order harmonic of a global magneto-elastic axial (torsional) oscillation, in which the crust and core oscillate together, coupled with the strong magnetic field \citep{glampedakis2006,andersson2009,steiner2009,vanhoven2011,vanhoven2012,colaiuda2011,colaiuda2012,gabler2012,gabler2013,passamonti2013b,passamonti2014,glampedakis2014}.

With two detections from two different data sets and two different sources, there is now an acute need for theoretical thought on whether small bursts could indeed excite QPOs either via crust fractures or explosive reconnection in the magnetosphere. At present, modelling work focusses largely on the QPOs observed from giant flares and the associated energetics. However, it is not clear whether the same processes can excite the same crustal shear and core modes postulated for giant flare QPOs in bursts that are of the order of $10^3$ times shorter and up to $\sim 10^{9}$ times smaller in energy compared to the giant flares.

There is no significant detection of a QPO in any of the individual bursts. At the same time, the \rxte\ data set illustrates a limiting case for the applicability of Fourier methodology: the combination of a complex burst morphology, with relevant timescales that can exceed $1000 \, \hz$, low source counts as well as a low background 
count rate render the basic assumption of many Fourier methods, including the methods presented in \citet{huppenkothen2013}, invalid. Because there is little background and a great amount of broadband source power even at high frequencies, frequencies are no longer independent of each other, and no longer distributed following the standard $\chi^2_2$ distribution invoked for periodograms of photon counting data. As we have shown, the distributions of powers are strongly shifted towards higher powers and are much broader than the expected distributions, leading to an increased probability of false-positive detections. In this case, simulations of light curves instead of periodograms, such as those introduced in Section \ref{sec:dnest}, offer a valid alternative that properly accounts for the changes in the periodogram shape. 

This problem is not necessarily limited to \rxte\ or to magnetar bursts: any instrument with low source counts and low background count rates (e.g. \textit{Swift}) will lead to similar effects in the periodograms of fast transient events with complex morphology (e.g. gamma-ray bursts). Harnessing the power of the light curve models described above for QPO searches, in combination with high-quality data from instruments such as \fermi/GBM, enables us to search for weak QPOs in transient events with an unprecedented sensitivity, be they magnetar bursts, GRBs or solar flares.

\acknowledgments
D.H. and ALW acknowledge support from a Netherlands Organization for Scientific Research (NWO) Vidi Fellowship (PI A. Watts).  

\bibliography{sgr_bursts_references}
\bibliographystyle{apj}

\end{document}